\documentclass[journal]{IEEEtran}

\ifCLASSINFOpdf
\else
\fi
\usepackage{cite}
\usepackage{amsmath,amssymb,amsfonts}
\usepackage{amsthm}
\usepackage{bm}
\usepackage{algorithmic}
\usepackage{graphicx}
\usepackage{textcomp}
\usepackage{xcolor}
\usepackage{booktabs}
\usepackage{algorithm}
\usepackage{algorithmic}
\usepackage{array}
\usepackage{multirow}
\usepackage{stfloats}
\usepackage[colorlinks,
            linkcolor=blue,       
            anchorcolor=blue,  
            citecolor=blue,        
            ]{hyperref}
\usepackage{tabularx}

\theoremstyle{definition}

\newtheorem{theorem}{Theorem}
\newtheorem{lemma}{Lemma}
\allowdisplaybreaks      

\begin{document}

\title{Conditional Diffusion Model with OOD Mitigation as High-Dimensional Offline Resource Allocation Planner in Clustered Ad Hoc Networks}

\author{Kechen Meng, Sinuo Zhang, Rongpeng Li, Chan Wang, Ming Lei, and Zhifeng Zhao
    \thanks{K. Meng, S. Zhang, R. Li, C. Wang, and M. Lei are with the College of Information Science and Electronic Engineering, Zhejiang University (email: \{mengkechen, 22431100, lirongpeng, 0617464, lm1029\}@zju.edu.cn).}
    \thanks{Z. Zhao is with Zhejiang Lab as well as the College of Information Science and Electronic Engineering, Zhejiang University (email: zhaozf@zhejianglab.com).}
}

%



\maketitle

\begin{abstract}
Due to network delays and scalability limitations, clustered ad hoc networks widely adopt Reinforcement Learning (RL) for on-demand resource allocation. Albeit its demonstrated agility, traditional Model-Free RL (MFRL) solutions struggle to tackle the huge action space, which generally explodes exponentially along with the number of resource allocation units, enduring low sampling efficiency and high interaction cost. In contrast to MFRL, Model-Based RL (MBRL) offers an alternative solution to boost sample efficiency and stabilize the training by explicitly leveraging a learned environment model. However, establishing an accurate dynamic model for complex and noisy environments necessitates a careful balance between model accuracy and computational complexity \& stability. To address these issues, we propose a Conditional Diffusion Model Planner (CDMP) for high-dimensional offline resource allocation in clustered ad hoc networks. By leveraging the astonishing generative capability of Diffusion Models (DMs), our approach enables the accurate modeling of high-quality environmental dynamics while leveraging an inverse dynamics model to plan a superior policy. Beyond simply adopting DMs in offline RL, we further incorporate the CDMP algorithm with a theoretically guaranteed, uncertainty-aware penalty metric, which theoretically and empirically manifests itself in mitigating the Out-of-Distribution (OOD)-induced distribution shift issue underlying scarce training data. Extensive experiments also show that our model outperforms MFRL in average reward and Quality of Service (QoS) while demonstrating comparable performance to other MBRL algorithms. 
\end{abstract}


%
\IEEEpeerreviewmaketitle

\section{Introduction}

\IEEEPARstart{W}{ireless} ad hoc networks have been widely applied in military and civilian fields, due to its merits in flexibility, low cost, and robustness\cite{4518343}. In this context, Time Division Multiple Access (TDMA) and its multi-frequency variant MF-TDMA have laid the very foundation for resource management therein \cite{9866568}. Moreover, the complexity of the topology and the locality of node observations in ad hoc networks significantly constrain the system's stability and reliability. To overcome these challenges, clustering methods\cite{9700761,9815245} can be naturally engaged to enhance coordination and improve efficiency. Nonetheless, the dynamic nature of service arrivals and departures often leads to underutilization in static resource allocation solutions, highlighting the need for a dynamic solution to adapt to underlying variations. Several algorithms have been proposed to address this challenge \cite{7148429,4803842}. Prominently, the success of Reinforcement Learning (RL) \cite{magnetic,MBAD} motivates the emergence of smart resource allocation methods\cite{9866568}. Typically, Model-Free RL (MFRL), such as DQN\cite{mnih2015human}, DDPG\cite{lillicrap2015continuous}, PPO\cite{schulman2017proximal}, are chosen to guide the learning of an optimal policy through direct interaction with the physical environment, which is costly and potentially unsafe in real-world applications. Moreover, MFRL suffers from low sample efficiency and unstable training in scenarios that require complex planning\cite{huang2020model}. Specifically, in the context of resource allocation in ad hoc networks with $N$ nodes, $M$ time slots, and $L$ channels, a typical setting of the action space could be $N^{M\times L}$, awfully scaling along with the increase of $L$, $M$ and $N$. To overcome these limitations, Model-Based RL (MBRL) has recently emerged as a promising approach to significantly improve the overall efficiency and effectiveness of RL algorithms\cite{10005026}.

Compared to MFRL, MBRL constructs an environment model from historical logs to simulate potential outcomes and generate synthetic trajectories, enabling the agent to fully leverage the experienced data while reducing trial-and-error costs\cite{huang2020model}. Consequently, decision-making can be framed as a planning problem, where the agent simulates future trajectories to select actions that achieve desired outcomes\cite{you2019advanced}. This approach improves the agent's ability to handle long-term dependencies and make informed decisions. However, since MBRL performance depends on accurate environment modeling, prediction errors, which naturally arise from the complexity and uncertainty of the dynamic environment, pose significant challenges. Additionally, traditional MBRL methods for planning problems rely on dynamic programming to stitch together sub-optimal reward-labeled trajectories in the training dataset to reconstruct optimal trajectories\cite{ajay2021opal,chua2018deep}. To enable dynamic programming, these approaches learn a value function that estimates the discounted sum of rewards from a given state. However, value function estimation faces significant instabilities due to function approximation, off-policy learning, and bootstrapping-collectively referred to as the ``deadly triad"\cite{712192}. These issues pose substantial barriers to scaling existing MBRL algorithms effectively.

On the other hand, since online policy training in ad hoc networks often faces high costs to interact with the environment, 
offline RL\cite{Lange2012} is favored 
to train agents only with pre-collected offline datasets while saving the cost of online interaction. 
However, relying solely on previously collected data often leads to poor policies that prefer Out-of-Distribution (OOD) actions with overestimated values, resulting in unsatisfactory performance\cite{Fujimoto2018OffPolicyDR}. This phenomenon is primarily attributed to the distributional shift \cite{Levine2020OfflineRL} between the learning and behavior policies, induced by the limited size and diversity of the training data. As a result, developing effective techniques to address this challenge has become a critical focus in the offline MBRL research community. Recent methods like MOPO\cite{yu2020mopo} and MOREC\cite{luo2023reward}, which incorporate a reward penalty based on an estimate of the uncertainty using model ensembles and learn reward-consistent dynamics through adversarial games respectively, partially mitigate this issue but still struggle to balance between model accuracy and computational complexity \& stability.

In this paper, to accommodate allocated resources to imbalanced traffic loads in clustered ad hoc networks, we propose a Conditional Diffusion Model Planner (CDMP) with offline training. Specially, in contrast to earlier deep generative methods like Energy-Based Models (EBMs)\cite{du2019implicit}, Variational Autoencoders (VAEs)\cite{kingma2013auto} and Generative Adversarial Networks (GANs)\cite{goodfellow2014generative}, Diffusion Models (DMs) exhibit superior capabilities in generating high-quality samples and demonstrate enhanced training stability, epitomizing the forefront of advancements in visual and language tasks\cite{brown2020language,rombach2022high}. Notably,
different from the DM-assisted MFRL \cite{10753523,liu2024generative,lillicrap2015continuous,10736570}, we first use a conditional DM to characterize environmental dynamics and guide planning towards trajectories with high cumulative future rewards. Afterward, we employ an inverse dynamics model to generate actions based on state transitions between adjacent time steps, thus benefiting the obtaining of a dynamic resource allocation policy. Therefore, our methodology works completely in an offline MBRL manner. Furthermore, to mitigate potential OOD issues in offline training, we develop CDMP-pen, a variant of CDMP, with a theoretically guaranteed penalty metric that can more effectively quantify the uncertainty of the estimated dynamics model. 
Compared to the existing literature, the contribution of the paper can be summarized as follows:
\begin{itemize}
\item To implement load-aware resource allocation in clustered ad hoc networks, we propose a sample-efficient MBRL solution (i.e., CDMP) that leverages dynamic programming to combine sub-optimal trajectories into an optimal one without relying on inaccurate value estimation. By modeling the environmental dynamics as a conditional DM, CDMP can optimize the resource scheduling strategy in MF-TDMA through offline training. Due to the complexity of modeling high-dimensional action sequences, we model the diffusion process between states only and then utilize an inverse dynamics model to predict actions. 
\item To mitigate OOD issues in the MBRL offline setting, in CDMP-pen, we quantify the risk arising from imperfect dynamics and appropriately balance that risk with return. By calibrating a ``smoothed distance to data'' metric to measure model uncertainty, we can guide the generation of high-confidence trajectories, ensuring that the decision variables remain consistently close to the offline dataset. Furthermore, we theoretically demonstrate that the smoothed distance to data serves as an upper bound for the true uncertainty based on the Lipschitz constant, thereby validating the soundness and effectiveness of the proposed metric.
\item We further illustrate our approach experimentally on high-fidelity OPNET simulations and validate the superiority of our proposed framework over other offline RL methods\cite{farag2018behavior,kumar2020conservative,Kostrikov2021OfflineRL,chen2021decision,janner2022planning}. In addition, we conduct extensive ablation studies to corroborate the robustness and reliability of our approach under varying conditions and parameter settings, further highlighting its practicality and adaptability.
\end{itemize}

The remainder of the paper is organized as follows. Section~\ref{II} briefly introduces the related works. In Section~\ref{III}, we introduce the preliminaries and formulate the system model. The details of our proposed framework are presented in Section~\ref{III}. In Section~\ref{IV}, we provide the results of extensive simulations and numerical analysis. Finally, the conclusion is summarized in Section~\ref{V}.

For convenience, we list the major notations of this paper in Table \ref{tab1}.

\begin{table}[tbp]
\caption{Main parameters and notations used in this paper.}
\begin{center}
\begin{tabular}{|c|m{6cm}|}
\hline
\toprule 
\textbf{Notations} & \textbf{Definition} \\
\midrule 
$N$ & Number of nodes \\
\hline
$M$ & Number of time slots in a frame \\
\hline
$L$ & Number of channels \\
\hline
$K$ & Diffusion steps \\
\hline
$H$ & Planning horizon \\
\hline
$N_\text{epoch}$ & Number of training epochs \\
\hline
$N_\text{step}$ & Number of training steps per epoch \\
\hline
$\epsilon_{\theta}$ & Diffusion noise model \\
\hline
$f_{\phi}$ & Inverse dynamics model \\
\hline
$\bm{x}_k$ & The intermediate latent variable at diffusion step $k$ \\
\hline
$\bm{y}$ & The conditioning information in conditional diffusion model \\
\hline
$s_t$, $a_t$, $\mathcal{R}_{t}$ & State, action and reward of the agent at $t$ \\
\hline
$\rm{Return}(\tau)$ & The return of the trajectory $\tau$ \\
\hline
$\widetilde{s}_t$ & The perturbed state \\
\hline
$\hat{s}_t$ & The reconstructed state \\
\hline
$s_c$ & The closest data-point in the state dataset \\
\hline
$d_\sigma(\widetilde{s};\mathcal{D}_{\rm{stat}})^2$ & The smoothed distance to data \\
\hline
$d(s;\mathcal{D}_{\rm{stat}})^2$ & The standard squared distance \\
\hline
$e(s)$ & The true uncertainty between the data distribution and the surrogate distribution \\
\hline
${\rm gen}_{t}^{(i)}$ & Number of packets generated within $t$ \\
\hline
${\rm sen}_{t}^{(i)}$ & Number of packets sent within $t$ \\
\hline
$T_{t}^{(i)}$ & Length of packet transmission queue at $t$ \\
\hline
$T_{\max,t}^{(i)}$ & Maximum length of packet transmission queue within $t$ \\
\hline
$n_{lm}$ & Node $n$ allocated to time slot $m$ of channel $l$ \\
\hline
$d_{t}^{(i)}$ & Time delay of node $i$ at $t$ \\
\hline
$u_{t}^{(i)}$ & Throughput of node $i$ at $t$ \\
\hline
$l_{t}$ & System packet loss rate at $t$ \\
\hline
$\lambda, \eta$ & Factors balancing the metrics in weighted reward function \\
\hline
$\omega$ & Conditional guidance scale \\
\hline
$\xi$ & Low-temperature sampling factor\\
\hline
$\beta$ & Parameter controlling the probability of removing conditional information \\
\hline
$\zeta$ & Parameter controlling the weight of OOD penalty term \\
\bottomrule 
\end{tabular}
\label{tab1}
\end{center}
\vspace{-0.5cm}
\end{table}


\section{Related Works}\label{II}
\subsection{Resource Schedule in Ad Hoc Networks}
In the TDMA Multiple Access Channel (MAC) protocol, efficient time slot scheduling algorithms are essential for managing channel access and optimizing resource utilization\cite{Zhu1998fivephase}. Classical methods, such as DTSS\cite{7148429}, a bandwidth-efficient dynamic TDMA slot scheduling scheme to accommodate unpredictable network traffic, and DDMC-TDMA\cite{jabandzic2021dynamic}, a topology-agnostic protocol designed for large-scale, high-density ad hoc networks, represent significant developments in this field. However, these approaches rely on predefined rules and static strategies, limiting adaptability in complex, real-time scenarios with high loads or bursty traffic. To overcome these limitations, MFRL-based methods have been introduced to enable more flexible and responsive scheduling. For instance, \cite{wi2020delay} proposes an actor-critic algorithm to optimize TDMA scheduling and minimize weighted end-to-end delay, a metric reflecting tactical traffic criticality. Similarly, \cite{Chilukuri2021Deadline} utilizes node features for state representation and applies diverse heuristics across time slots, enabling adaptive decisions that account for long-term deadline adherence. While these RL-based methods leverage real-time feedback to dynamically adjust scheduling strategies and enhance adaptability over traditional rule-based approaches, they still encounter challenges in long-term decision-making, convergence to local optima, and training instability. Moreover, their online RL nature incurs substantial interaction costs. To address these issues, we incorporate DMs with offline RL for MF-TDMA resource allocation, which better captures long-term dependencies and enables global optimization, thereby 
providing a more robust solution for high-dimensional resource management.

\subsection{Diffusion Models for Resource Allocation in Communication Systems}
Although prior studies have exemplified the expanding role of DMs in tackling complex problems beyond their traditional generative contexts\cite{xian2023chaineddiffuser,he2023diffusion}, their use for managing wireless networks and optimizing resource allocation strategy remains underexplored \cite{10529221}. These networks exhibit complex dynamics and highly mobility of wireless devices\cite{9798257}. Additionally, the impact of noise and interference in wireless channels, which affects the generalization of RL policies, is often overlooked \cite{10515203}. Recently, several studies have concentrated on harnessing the capabilities of DMs in data augmentation, latent space representation, and knowledge transfer, to effectively deal with numerous optimization problems of intelligent networks. For example, \cite{10753523} proposes a DDPM-based framework to train the resource allocation algorithm that generates optimal block lengths conditioned on the channel state information. D3PG \cite{liu2024generative} proposes an access mechanism that integrates DMs with DDPG \cite{lillicrap2015continuous} framework to jointly adjust the contention window and the aggregation frame length in Wi-Fi networks, while \cite{10736570} extends to the multi-agent scenario that incorporates DMs to determine the optimal DNN partitioning and task offloading decisions. However, these works primarily focus on the MFRL framework and utilize DMs as the agent’s policy, which struggle with high-dimensional actions and environmental uncertainties. In this paper, we adopt the planning framework and utilize DMs to capture the highly complex dynamics of the environment, thus enabling the MBRL agent to better leverage temporal correlations and enhancing the overall decision-making process.

\subsection{Planning with Model-Based Reinforcement Learning}
To tackle the sampling inefficiency and communication overhead in MFRL, MBRL naturally emerges as an alternative solution. By leveraging a learned dynamics model, agents can predict future trajectories and achieve higher returns through planning, which is a fundamental research focus in MBRL. However, MBRL algorithms often struggle to scale effectively as the complexity of RL tasks increases. Moreover, modeling errors in dynamics tend to accumulate over time, significantly limiting the practical applications of MBRL approaches. Focusing on these challenges, PETS\cite{chua2018deep} combines uncertainty-aware dynamics models with sampling-based uncertainty propagation using probabilistic ensembles. POPLIN\cite{Wang2020Exploring} formulates action planning at each time-step as an optimization problem, and argues that optimization in the policy network’s parameter space will be more efficient. On the other hand, PILCO\cite{pilco2011} explicitly marginalizes aleatoric and epistemic uncertainty in the dynamics model to directly optimize the surrogate expected reward for long-term planning. Although these works mitigate some of the aforementioned problems, they either rely on model ensembles for uncertainty estimation, which becomes computationally prohibitive in high-dimensional environments, or require value function fitting for dynamic programming, resulting in severe cumulative errors and training instability. Additionally, conventional MBRL approaches typically rely on fully connected networks parameterizing diagonal-covariance Gaussian distributions as the environment model, which struggle to accurately represent complex multi-modal distributions. Inspired by recent advancements \cite{NEURIPS2021_099fe6b0,chen2021decision}, we reinterpret the planning problem in MBRL as a sequence generation task, thereby alleviating the computational burden while fully leveraging the expressiveness and robustness of DMs.

\subsection{Addressing Out-of-Distribution in Offline RL}
Although offline RL avoids the risks and time-consuming nature of real-world interactions, it faces large extrapolation errors when the Q-function is evaluated on OOD actions, which can lead to unstable learning and potential divergence. To address these challenges, recent offline RL algorithms have introduced conservative value function estimation to enhance the reliability of model-based rollouts. For instance, COMBO\cite{yu2021combo} learns a conservative critic by penalizing the value function for state-action pairs outside the support of the offline dataset, while OSR\cite{jiang2023recovering} further incorporates Conservative Q-Learning (CQL) \cite{kumar2020conservative} with an estimated inverse dynamics model to implement state recovery. Another class of approaches focuses on optimizing a lower bound of policy performance, which is constructed using uncertainty estimates of the learned model. MOPO\cite{yu2020mopo} penalizes rewards through the aleatoric uncertainty in next-state predictions, and MOBILE\cite{Sun2023ModelBellmanIF} introduces an uncertainty quantifier to estimate the Bellman errors induced by the learned dynamics through the inconsistency of Bellman estimations under different learned models. Drawing from these methodologies, we integrate uncertainty-based penalization to avoid unreliable model predictions, aiming to more effectively balance between model accuracy and computational complexity \& stability.

\section{Preliminaries \& System Model}\label{III}
In this section, we briefly introduce the framework of MBRL and some fundamentals of DMs. On this basis, we highlight how to apply DMs to tackle the sequential decision-making issue for resource allocation in clustered ad hoc networks.

\subsection{Preliminaries}
\paragraph{Model-Based Reinforcement Learning} We formulate the sequential decision-making problem as a discounted Markov Decision Process (MDP), defined by the tuple $\left\langle \mathcal{S}, \mathcal{A}, \mathcal{T}(s_{t+1}|s_t,a_t),\mathcal{R}(s_t, a_t), \gamma \right\rangle$\cite{janner2022planning}. In this formulation, $\mathcal{S}$ represents the set of possible states and $\mathcal{A}$ is the set of actions the agent can take. The transition function $\mathcal{T}$ describes the probability of transitioning from state $s_t$ to $s_{t+1}$ after executing an action $a_t$. The reward function $\mathcal{R}$ gives the immediate reward the agent receives after taking action $a_t$ in state $s_t$, and the discount factor $\gamma \in \left[0,1\right)$ balances the importance of future rewards, with lower values favoring immediate rewards. The agent follows a stochastic policy $\pi$, mapping states to actions. In RL, the standard goal of the agent is to learn a policy that maximizes the expected cumulative discounted sum of rewards in trajectory $\tau := \{s_0,a_0,s_1,a_1,\cdots\}$ as
\begin{equation}
\pi^* := \mathop{\arg\max}\limits_{\pi} \mathbb{E}_{\tau\sim p_{\pi}}\left[\sum\nolimits_{t\ge0}\gamma^{t}\mathcal{R}(s_t,a_t)\right].
\end{equation}
In MBRL, the agent first estimates the dynamics model from collected environmental transition samples. Afterward, based on the learned dynamics model, the agent predicts subsequent actions and executes a planning phase to optimize its policy\cite{janner2019trust}. This characteristic significantly reduces the physical interactions between the agent and the environment, which results in higher data efficiency and lower sampling costs\cite{huang2020model}.

\paragraph{Diffusion Probabilistic Models} We try to use $p_{\theta}(\bm{x})$ to approximate the data distribution $q(\bm{x})$ from the dataset $\mathcal{D}:=\left\{\bm{x}^i\right\}_{0\le i \le M}$ based on DMs, which can be achieved by minimizing the KL divergence between both distributions\cite{gen2022taylor}. By the definition of divergence, this objective can be reformulated as maximizing the log-likelihood as
\begin{align}
\theta^* &:= \mathop{\arg\min}\limits_{\theta} D_\text{KL}(q(\bm{x}) \parallel p_{\theta}(\bm{x}))\nonumber \\
         &= \mathop{\arg\min}\limits_{\theta} \mathbb{E}_{\bm{x} \sim q(\bm{x})} \left[\log q(\bm{x}) - \log p_{\theta}(\bm{x})\right]\nonumber \\
         &= \mathop{\arg\max}\limits_{\theta} \mathbb{E}_{\bm{x} \sim q(\bm{x})} \left[\log p_{\theta}(\bm{x})\right]         \label{eq:log-likelihood}\\
         &\approx \mathop{\arg\max}\limits_{\theta} \sum\nolimits_{i=1}^{M} \log p_{\theta}(\bm{x}^i).\nonumber 
\end{align}
As a powerful class of generative models, DMs transform the objective above into maximizing the Evidence Lower BOund (ELBO) of a Hierarchical Variational Auto-Encoder (HVAE)\cite{luo2022understanding}. By modeling the data generation process through reversing a forward noising process, DMs effectively perform a denoising process. For each data point $\bm{x}_0 \sim q(\bm{x})$ from the dataset $\mathcal{D}$, the predefined forward noising process can be modeled as a discrete Markov chain $\bm{x}_{0:K}$ such that $q(\bm{x}_k|\bm{x}_{k-1}) = \mathcal{N}(\bm{x}_k|\sqrt{\alpha_k}\bm{x}_{k-1},(1-\alpha_k)\bm{I})$. Here, $\mathcal{N}(\mu, \Sigma)$ denotes a Gaussian distribution with mean $\mu$ and variance $\Sigma$, and $0<\alpha_K<\cdots<\alpha_0<1 \in \mathbb{R}$ are hyperparameters controlling the variance schedule. For appropriately chosen $\alpha_k$ and a sufficiently large $K$, we assume that $\bm{x}_K \sim \mathcal{N}(\textbf{0}, \bm{I})$. The trainable reverse process modeled as the variational reverse Markov chain is parameterized with $p_{\theta}(\bm{x}_{k-1}|\bm{x}_k) = \mathcal{N}(\bm{x}_{k-1}|\mu_{\theta}(\bm{x}_k,k),\Sigma(\bm{x}_k,k))$. This allows us to sample $\bm{x}_K$ from standard Gaussian noise and progressively denoise it until obtaining $\bm{x}_0$. The reverse process can be learned by optimizing a simplified surrogate loss as
\begin{equation}
\mathcal{L}(\theta):=\mathbb{E}_{\substack{k\sim\left[1,K\right]\\\bm{x}_0\sim q\\\epsilon\sim\mathcal{N}(\textbf{0}, \bm{I})}}\left[{\parallel\epsilon-\epsilon_{\theta}(\bm{x}_k,k)\parallel}^2\right].\label{eq:reverse_loss}
\end{equation}
Here, the deep neural network $\epsilon_{\theta}$ parameterized by $\theta$ is trained to predict the noise $\epsilon \sim \mathcal{N}(\textbf{0}, \bm{I})$, which is added to the dataset sample $\bm{x}_0$ to produce $\bm{x}_k$ as
\begin{equation}
\bm{x}_k:=\sqrt{\overline{\alpha}_k}\bm{x}_0+\sqrt{1-\overline{\alpha}_k}\epsilon,\label{eq:reconstruct}
\end{equation}
where $\overline{\alpha}_k:=\prod_{s=1}^{k}\alpha_s$. This is equivalent to predicting the mean of $p_{\theta}(\bm{x}_{k-1}|\bm{x}_k)$, since $\mu_{\theta}(\bm{x}_k,k)$ can be calculated as a function of $\epsilon_{\theta}(\bm{x}_k,k)$\cite{ho2020denoising} as
\begin{equation}
\label{eq:denoising_mean}
\mu_{\theta}(\bm{x}_k,k)=\frac{1}{\sqrt{\alpha_k}}\bm{x}_k-\frac{1-\alpha_k}{\sqrt{1-\overline{\alpha}_k}\sqrt{\alpha_k}}\epsilon_{\theta}(\bm{x}_k,k).
\end{equation}

\subsection{System Model}
\begin{figure}[tbp]
\centerline{\includegraphics[width=0.5\textwidth]{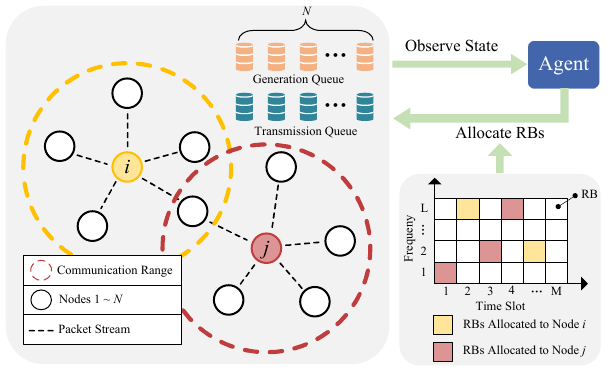}}
\caption{Ad hoc network topology under MF-TDMA MAC protocol.}
\label{fig1}
\vspace{-0.2cm}
\end{figure}

We consider an adaptive, resource-constrained clustered ad hoc network scenario with $N$ nodes, as illustrated in Fig.~\ref{fig1}. In this setup, each node communicates directly with its one-hop neighbors using wireless transceivers, while communication with remote nodes requires multi-hop relaying. Every node is equipped with packet queues to cache data packets. If a sent packet remains unacknowledged, it will be appended to the end of the cache queue of forwarding nodes in case. We assume the traffic loads therein are prioritized, where a portion of nodes generate packets at high rates while others at relatively slower rates. However, the extent of imbalance and variations in traffic loads are unknown in advance. Specifically, we employ MF-TDMA \cite{9866568} as the MAC layer protocol. In this setting, the practical implementation in one frame encompasses two stages (i.e., network management and traffic transmission).  
In the first stage, nodes perform time synchronization and interact with the agent to determine the allocation scheme. The second stage can be divided into $ML$ Resource Blocks (RBs), where the nodes start to transmit packets within allocated RBs efficiently.

To optimize resource utilization and minimize collision probability, it is essential to leverage a dynamic resource allocation mechanism to allocate an appropriate number of RBs to each node according to individual characteristics. Therefore, we formulate the resource allocation problem as an MDP, and define the decision interval as one frame. Specifically, at each time-step (i.e., one frame) $t$, the agent observes the packet queue length of all nodes in the system. The state of node $i$ is defined as $s_{t}^{(i)}:=\left [  {\rm gen}_{t}^{(i)}, {\rm sen}_{t}^{(i)},T_{t}^{(i)},T_{\max,t}^{(i)} \right ]$, where ${\rm gen}_{t}^{(i)}$ and ${\rm sen}_{t}^{(i)}$ denote the numbers of packets generated and sent within the frame $t$, while $T_{t}^{(i)}$ and $T_{\max,t}^{(i)}$ represent the real-time transmission queue length at $t$ and maximum length within frame $t$, respectively. The overall state is the concatenation of the states of all nodes, i.e., $s_{t}:=\left [ s_{t}^{(1)},s_{t}^{(2)} ,\cdots, s_{t}^{(N)} \right ]$. Based on the observed state $s_{t}$, the agent takes action $a_{t}:=\left [ n_{lm}  \right ] \in \mathbb{Z}^{L\times M}$ to allocate RBs, where $n_{lm}$ denotes the node $n$ assigned to time slot $m$ of channel $l$. To improve the overall system Quality of Service (QoS), we define the reward function as 
\begin{equation}
\mathcal{R}_{t} :=- \sum\nolimits_{i=1}^{N} d_{t}^{(i)} , 
\label{eq:reward}
\end{equation}
where $d_{t}^{(i)}$ indicates the time delay for packets received by node $i$. Our experiment experience shows that by minimizing the average delay, the communication system can achieve improved throughput and reduced packet loss rate, enhancing the overall QoS. Meanwhile, it avoids the cumbersome tuning of multi-objective optimization weights. In this paper, we resort to a conditional DM-based high-dimensional resource allocation planner for maximizing the cumulative reward function.

\section{Resource Allocation With CDMP \& CDMP-pen}\label{IV}
\begin{figure*}[tbp]
\setlength{\abovecaptionskip}{0cm} 
\setlength{\belowcaptionskip}{-0.5cm} 
\centering
\includegraphics[width=\textwidth]{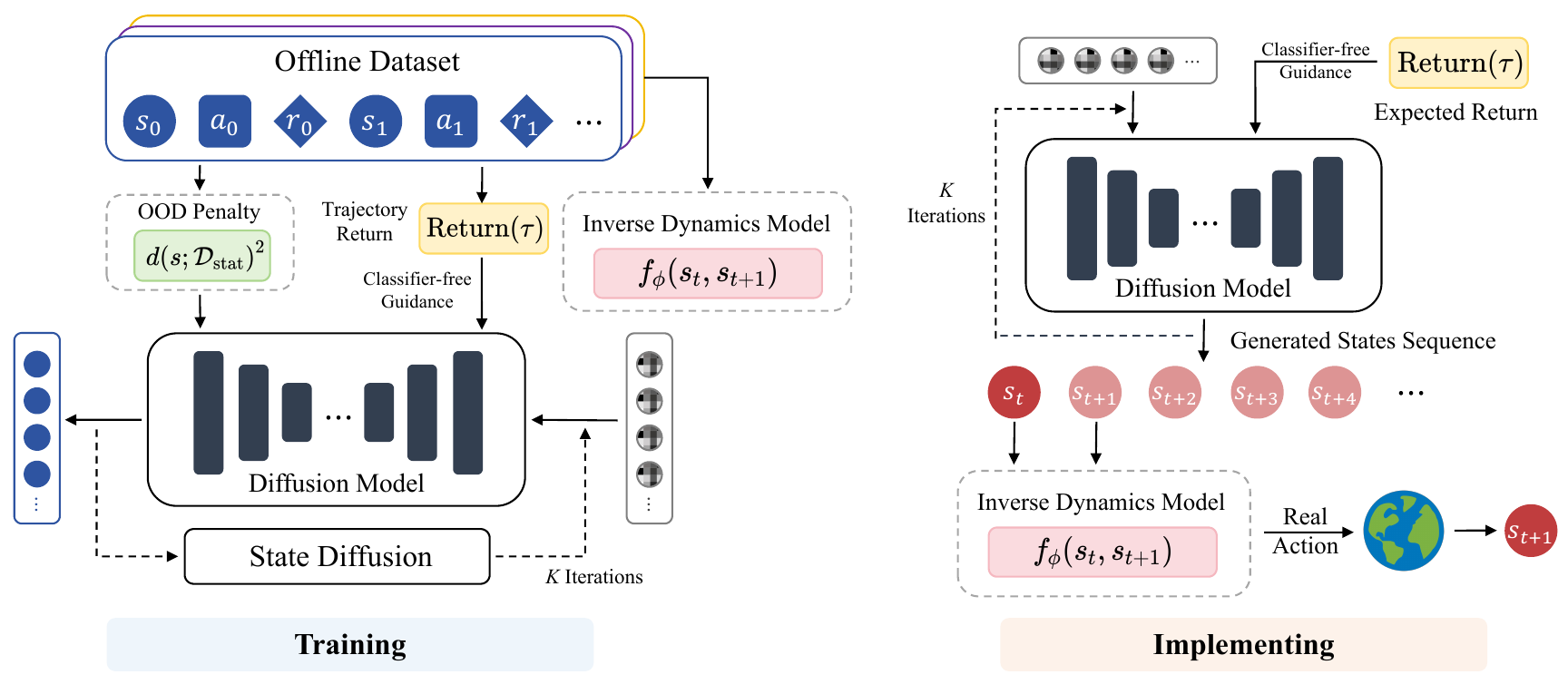}
\caption{Training and implementation of CDMP \& CDMP-pen. }
\label{fig2}
\vspace{-0.2cm}
\end{figure*}

In this section, we discuss how to apply conditional DMs to guide the generation of trajectories with high confidence toward superior cumulative returns. We elaborate on the CDMP framework for policy optimization through offline training. Furthermore, we develop CDMP-pen with uncertainty estimation to effectively combat the OOD problem. The entire procedure for training and implementation is illustrated in Fig.~\ref{fig2}.  
\subsection{Planning over Decision Sequences}

In the planning phase of MBRL, it usually needs to simulate or explore different sequences of actions and states, thus evaluating outcomes and evolving decisions from the perspective of a longer horizon. 
Correspondingly, we formulate the state and/or action trajectory modeling as a DM-based conditional generative problem\cite{ajay2022conditional}, that is
\begin{equation}
\max\limits_{\theta}\mathbb{E}_{\tau \sim \mathcal{D}}\left[\log p_{\theta}(\bm{x}_0(\tau)|\bm{y}(\tau))\right].
\label{equa}
\end{equation}
In other words, after characterizing the distribution $p_{\theta}$ of the trajectories sampled from an MDP, we can conveniently generate portions of a trajectory $\bm{x}_0(\tau)$ from information $\bm{y}(\tau)$ about it. Here, in the context of RB allocation in clustered ad hoc networks, the discreteness of actions, with higher frequency and less smooth characteristics, adds to the difficulty of predicting and modeling. Therefore, we define $\bm{x}_k(\tau)$ as a noisy sequence of states with a length $H$, namely
\begin{equation}
\bm{x}_k(\tau):={(s_t, s_{t+1}, \cdots, s_{t+H-1})}_k.
\end{equation}
Meanwhile, to guide the denoising process toward generating trajectories with high cumulative rewards, $\bm{y}(\tau)$ could refer to the return of the trajectory $\rm{Return}(\tau)$, that is
\begin{equation}
\label{eq:return}
\bm{y}(\tau):= {\rm{Return}}(\tau):=\sum\nolimits_{t'=t}^{t+H-1}{\gamma}^{t'-t}\mathcal{R}_{t'}.
\end{equation}

Next, contingent on a learned inverse dynamics model $f_{\phi}(s_t, s_{t+1})$ \cite{allen2021learning}, the training of which will be given in Section \ref{sec:training}, for any timestep $t$ in $\bm{x}_0(\tau)$, the policy is inferred by estimating the action that leads to the transition from state $s_t$ to $s_{t+1}$, namely
\begin{equation}
a_t:=f_{\phi}(s_t, s_{t+1}).
\end{equation}
Therefore, we can accomplish trajectory-based planning to implement policy optimization.

\subsection{Classifier-free Guidance-based Training}
\label{sec:training}
In order to guide trajectory generation for more efficient planning and decision-making, conditional information is incorporated into the diffusion process to ensure that the generated trajectories align with the desired attributes. One possible solution is classifier-based guidance\cite{dhariwal2021diffusion}, which trains a classifier independently to predict condition $\bm{y}(\tau)$ from noisy trajectories $\bm{x}_k(\tau)$, and the gradients from the classifier are used to guide the sampling process of an unconditional diffusion model. However, this method introduces additional computational overhead and heightened sensitivity to classifier performance, as training a classifier is equivalent to estimating a Q-function to predict the discounted cumulative reward from timestep $t$.  

Therefore, instead of training a separate classifier model, we utilize classifier-free guidance\cite{ho2022classifier}, which mixes the score estimates of a conditional diffusion model and a jointly trained unconditional diffusion model as $((1-\beta)\bm{y}(\tau)+\beta \varnothing,k)$. Here, the parameter $\beta$, drawn from a Bernoulli distribution, controls the interpolation between the conditional and unconditional guidance. In other words, we use a unified neural network to parameterize both models, whereas for the unconditional model, we can simply input a null token $\varnothing$. On this basis, consistent with Eq. \eqref{eq:reverse_loss}, we train an unconditional denoising diffusion model $p_{\theta}(\bm{x}_{k-1}(\tau)|\bm{x}_k(\tau))$ parameterized through a noise estimator $\epsilon_{\theta}(\bm{x}_k(\tau), k)$ together with the conditional model $p_{\theta}(\bm{x}_{k-1}(\tau)|\bm{x}_k(\tau), \bm{y}(\tau))$ parameterized through $\epsilon_{\theta}(\bm{x}_k(\tau), \bm{y}(\tau), k)$. 

We simultaneously train the reverse diffusion process $p_{\theta}$, which can be reasonably learned from the noise model $\epsilon_{\theta}$ Eq. \eqref{eq:denoising_mean}, and the inverse dynamics model $f_{\phi}$ in a supervised manner. Given an offline dataset $\mathcal{D}$ that consists of state-action trajectories labeled with rewards, we train with the reverse diffusion loss and the inverse dynamics loss:
\begin{align}
\label{eq:CDMP}
&\mathcal{L}_{\rm{CDMP}}(\theta,\phi):=\mathbb{E}_{(s,a,s')\in \mathcal{D}}\left[{\parallel a-f_{\phi}(s,s')\parallel}^2\right]\\
&\quad +\mathbb{E}_{k,\tau \in \mathcal{D},\beta}\left[{\parallel\epsilon-\epsilon_{\theta}(\bm{x}_k(\tau),(1-\beta)\bm{y}(\tau)+\beta \varnothing,k)\parallel}^2\right].\nonumber
\end{align}

\subsection{OOD Penalty Based on Distance to Data}
Although offline MBRL leverages pre-collected data, which is often more cost-effective compared to online interactions, it inherently lacks real-time data exchange. This limitation hinders the model's adaptability to unfamiliar states or dynamic changes in the environment, ultimately leading to the OOD problem. 
Recalling the derivations in Eq. \eqref{eq:log-likelihood}, we try to remedy this by introducing a constrained conservative objective for individual states as
\begin{equation}
\label{eq:cons_obj}
\min\limits_{\theta} \sum\nolimits_{t'=t+1}^{t+H-1}e(s_{t'}),
\end{equation}
where $e(s_{t'}):=\| \log p_{\theta}(s_{t'})-\log q(s_{t'})\| _2$ is defined as the true uncertainty between the true distribution $q$ and the surrogate distribution $p_{\theta}$. 
Eq. \eqref{eq:cons_obj} restricts the generated states of the trajectory to closely match the training distribution, thus making the generation more reliable. 

\begin{figure}[tbp]
\centerline{\includegraphics[width=0.5\textwidth]{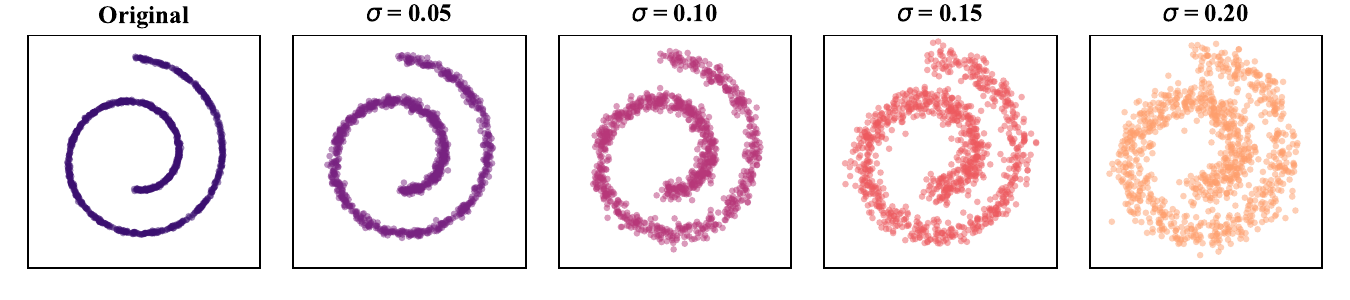}}
\caption{Illustration of the perturbed two-dimensional swiss roll distribution.}
\label{swiss roll}
\vspace{-0.2cm}
\end{figure}

However, directly computing likelihoods for high-dimensional data poses significant challenges. Meanwhile, the scarcity of the dataset adds to the difficulty. Consistent with prior works\cite{Laskey2017DARTNI,Gen2019song}, we first perturb data with random Gaussian noise 
to smooth the empirical distribution and construct a more comprehensive state space that includes meaningful OOD states\cite{SDC2022zhang}. Consider the state dataset $\mathcal{D}_{\rm{stat}}:=\left\{s^i\right\}_{0\le i \le M'}$ and its corresponding empirical distribution $q(s;\mathcal{D}_{\rm{stat}})$, we define the perturbed empirical distribution \cite{Gen2019song} as
\begin{align}
q_\sigma(\widetilde{s}; \mathcal{D}_{\rm{stat}}) &:= \int q(s; \mathcal{D}_{\rm{stat}}) \mathcal{N}(\widetilde{s}; s, \sigma^2 \mathbf{I}) ds \nonumber\\
&= \frac{1}{M'} \sum_i \mathcal{N}(\widetilde{s^i}; s^i, \sigma^2 \mathbf{I}),
\end{align}
where $\sigma$ is the noise level controlling the degree of perturbation applied to the original data. As demonstrated by the two-dimensional Swiss roll distribution in Fig.~\ref{swiss roll}, an increase in $\sigma$ results in a progressively smoother distribution. Afterward, we try to reformulate Eq. \eqref{eq:cons_obj} by introducing an uncertainty penalty term based on the smoothed distance to data \cite{Suh2023FightingUW}. Taking an arbitrary point in the smoothed state space $\widetilde{s} \in \mathbb{R}^n$, we formally define the smoothed distance to data as
\begin{align}
 d_\sigma(\widetilde{s};\! \mathcal{D}_{\rm{stat}})^2\! &:= {\rm{Softmin}}_\sigma \frac{1}{2} \|\widetilde{s} - s^i\|_2^2 + C_1 \nonumber\\
 &\!=\! \textstyle{-\sigma^2 \log \! \left[ \sum_i \exp\! \left( -\frac{1}{2\sigma^2} \|\widetilde{s} \! - \! s^i\|_2^2 \right) \right]\! +\! C_1,}\label{eq:smoothed distance to data}
\end{align}
where $C_1$ is the constant to ensure positiveness of $d_\sigma(\widetilde{s}; \mathcal{D}_{\rm{stat}})^2$. Different from \cite{Suh2023FightingUW}, which directly optimizes actions based on policy gradient methods, we apply this ``smoothed distance to data'' metric to constrain the generated states for conservative planning. More precisely, the rationality of using the smoothed distance to data as an uncertainty metric can be explained in the following lemma.
\begin{lemma}\label{lemma:equivalence}
    The negative log-likelihood of the perturbed empirical distribution $q_\sigma(\widetilde{s}; \mathcal{D}_{\rm{stat}})$ is equivalent to the smoothed distance to data by the noise level $\sigma$, up to some constant that does not depend on $s$,
    \begin{equation}
        -\sigma^2 \log q_\sigma(\widetilde{s}; \mathcal{D}_{\rm{stat}})= d_\sigma(\widetilde{s}; \mathcal{D}_{\rm{stat}})^2+C(M', n, \sigma),
    \end{equation}
    where we define $C(M', n, \sigma) := \sigma^2 (\log M' +  n/2 \log(2\pi\sigma))$.
\end{lemma}
The proof of Lemma \ref{lemma:equivalence} is included in Appendix \ref{proof:lemma 1}. Intuitively, Lemma \ref{lemma:equivalence} explicitly links the likelihood and the ``smoothed distance to data'' metric. 
Unlike aforementioned uncertainty-penalty-based methods \cite{yu2020mopo,Sun2023ModelBellmanIF} that rely on model ensembles or require value function fitting, this metric offers higher computational efficiency and models uncertainty with reasonable accuracy. 

Next, we mathematically establish the relationship between the ``smoothed distance to data'' metric and $e(s)$ in Eq. \eqref{eq:cons_obj}, by analyzing how much the metric provides an upper bound on the true uncertainty using the Lipschitz constant \cite{Marsden1974ElementaryCA} of the model bias $L_e$.
\begin{theorem}\label{theorem:bound}
    Let $L_e$ be the local Lipschitz constant of the true error $e(s)$ valid over the state space $\mathcal{S}$. For any $s \in \mathcal{S}$, $e(s)$ can be bounded by
    \begin{equation}
    \label{eq:bound}
        e(s) \leq e(s_c) + \sqrt{2} L_e \sqrt{d_\sigma(s; \mathcal{D}_{\rm{stat}})^2+C_2},
    \end{equation}
    where $s_c := \mathop{\arg\min}\limits_{s^i \in \mathcal{D}_{\rm{stat}}}\frac{1}{2} {\parallel s - s^i \parallel}_2^2$, i.e., the closest data-point, and $C_2 := \sigma^2 \log M' - C_1$, with $C_1$ defined in Eq. \eqref{eq:smoothed distance to data}.
\end{theorem}
The proof of Theorem \ref{theorem:bound} is unveiled in Appendix \ref{proof:lemma 2}. We believe this offers benefits over directly maximizing log-likelihood, as the smoothed distance to data avoids numerical instability and does not require explicit density estimation. In general, it is difficult to obtain $L_e$ in the absence of more structured knowledge of the empirical distribution $q(s)$. However, it is possible to obtain confidence bounds on $L_e$ using statistical estimation with pairwise finite slopes $\|e(s^i) - e(s^j)\| / \|s^i - s^j\|$ within the dataset\cite[Ch. 3]{Coles2001AnIT}\cite{9387079}. 
Based on Theorem \ref{theorem:bound}, the ``smoothed distance to data'' metric can overestimate the true uncertainty $e(s)$, enabling the establishment of an error bound and serving as a safeguard against excessive deviation from the true data distribution. Moreover, leveraging the powerful distribution modeling capability of DMs, we can assume that the first term on the right side of Eq. \eqref{eq:bound} approaches zero. Therefore, we primarily focus on optimizing the second term.

Additionally, as the smoothing degree approaches zero, i.e., $\sigma \rightarrow 0$, the smoothed distance to data converges to the standard squared distance $d{(s;\mathcal{D}_{\rm{stat}})}^2:=\min_{s^i \in \mathcal{D}_{\rm{stat}}} \frac{1}{2} {\| s - s^i \|}_2^2$. 
To encourage the surrogate state distribution closer to the real noise-free data distribution, we employ this true minimum distance $d{(s;\mathcal{D}_{\rm{stat}})}^2$ as the uncertainty metric during practical training. As a byproduct, it also enhances the safety of long-term planning and promotes policy conservativeness. 
Moreover, we relax this minimum distance by the deviation between the model-generated states and the input states within data batches. According to Eq. \eqref{eq:reconstruct}, the states generated by the model can be directly obtained from the reconstructed trajectories as
\begin{equation}
{(\hat{s}_t, \hat{s}_{t+1}, \cdots, \hat{s}_{t+H-1})}\!:=\!\hat{\bm{x}}_0(\tau)\!:=\!\frac{\bm{x}_k(\tau)\! -\! \sqrt{1 - \overline{\alpha}_k} \epsilon_{\theta}}{\sqrt{\overline{\alpha}_k}}.
\end{equation}
Based on this, we reformulate the conservative objective in Eq. \eqref{eq:cons_obj} as an OOD penalty term, namely
\begin{equation}
\mathcal{L}_{\rm{OOD}}(\theta):=\mathbb{E}_{k, \tau \in \mathcal{D}}\left[\sum\nolimits_{t'=t+1}^{t+H-1}{\parallel s_{t'}-\hat{s}_{t'}\parallel}^2\right],
\end{equation}
where $s_{t'}$ represents the state within the batch from the offline dataset. 
Since this relaxed distance is guaranteed to be greater than or equal to the true minimum distance, we can theoretically prove that this metric still bounds the true model error $e(s)$. Ultimately, we formulate the penalized loss as
\begin{equation}
\mathcal{L}_{\rm{CDMP-pen}}(\theta,\phi):=\mathcal{L}_{\rm{CDMP}}(\theta,\phi) + \zeta \mathcal{L}_{\rm{OOD}}(\theta),\label{eq:CDMP-pen loss}
\end{equation}
where $\zeta$ controls the weight of OOD penalty term. The overall offline training procedure of CDMP \& CDMP-pen is summarized in Algorithm ~\ref{training}, while the implementation process of both frameworks will be detailed in Section \ref{sec:implementing}.

\begin{algorithm}[!t]
    \caption{Offline training of CDMP \& CDMP-pen.}
    \label{training}
    \renewcommand{\algorithmicrequire}{\textbf{Input:}}
    \renewcommand{\algorithmicensure}{Initialize}
    \renewcommand{\algorithmiccomment}[1]{\hfill $\triangleright$ #1}
    
    \begin{algorithmic}[1]
        \REQUIRE Offline trajectory dataset $\mathcal{D}$ and state dataset $\mathcal{D}_{\rm{stat}}$, planning horizon $H$, diffusion steps $K$, conditional information removal probability $\beta$, OOD penalty weight $\zeta$
        \ENSURE Diffusion noise model $\epsilon_{\theta}$ and inverse dynamics model $f_{\phi}$ with random parameters $\theta$, $\phi$
        
        \FOR{each epoch}
          \FOR{each step}
            \STATE Sample a mini-batch $\mathcal{B}$ of $\bm{x}_0(\tau)$ in $\mathcal{D}$
            \STATE Predict actions $f_{\phi}(s,s')$ for each state transition pair in $\mathcal{B}$
            \STATE $\bm{y}(\tau) \gets {\rm{Return}}(\tau)$ \COMMENT{Obtain the condition information by Eq. \eqref{eq:return}}  
            \STATE $k \sim{\rm{Uniform}}(\{1,2,\cdots,K\})$ 
            \STATE $\epsilon \sim \mathcal{N}(\textbf{0}, \bm{I})$
            \STATE $\bm{x}_k(\tau)\gets\sqrt{\overline{\alpha}_k}\bm{x}_0(\tau)+\sqrt{1-\overline{\alpha}_k}\epsilon$ \COMMENT{Apply the forward diffusion process}
            \STATE Predict noise $\epsilon_{\theta}(\bm{x}_k(\tau),(1-\beta)\bm{y}(\tau)+\beta \varnothing,k)$
            \IF{training CDMP}
              \STATE Update the diffusion noise model $\epsilon_{\theta}$ and inverse dynamics model $f_{\phi}$ by Eq. ~\eqref{eq:CDMP}
            \ELSIF{training CDMP-pen} 
              \STATE ${(\hat{s}_t, \!\hat{s}_{t+1},\! \cdots\!, \!\hat{s}_{t+H-1}\!)}\! \gets \! \left( \bm{x}_k(\tau)\! -\! \sqrt{1\! -\! \overline{\alpha}_k} \epsilon_{\theta} \right)\! /\! \sqrt{\overline{\alpha}_k}
$\\ \COMMENT{Reconstruct trajectory}
              \STATE Update the diffusion noise model $\epsilon_{\theta}$ and inverse dynamics model $f_{\phi}$ by Eq. ~\eqref{eq:CDMP-pen loss}
            \ENDIF
            \ENDFOR
        \ENDFOR
    
    \end{algorithmic}
\end{algorithm}

\subsection{Implementation}
\label{sec:implementing}
We can implement the well-trained noise model and inverse dynamic model to generate trajectories for policy optimization. Formally, the trajectory generation process $p_{\theta}$ can be described as starting from Gaussian noise $\bm{x}_K(\tau) \sim \mathcal{N}(\textbf{0},\xi \bm{I})$ and diffuse $\bm{x}_k(\tau)$ into $\bm{x}_{k-1}(\tau)$ at each diffusion step $k$ with the perturbed noise as
\begin{align}
    &\hat{\epsilon}:=\epsilon_{\theta}(\bm{x}_k(\tau),\varnothing,k)+\nonumber\\
    &\quad \omega(\epsilon_{\theta}(\bm{x}_k(\tau),\bm{y}(\tau),k)-\epsilon_{\theta}(\bm{x}_k(\tau),\varnothing,k)),
\end{align}
where we manually set the condition $\bm{y}(\tau)$ as a high constant value during implementation, which serves as the conditional expected return, and the scalar $\omega$ adjusts the influence of the condition by amplifying and extracting the most relevant portions of trajectories that exhibit the desired attributes in the dataset. Meanwhile, to reduce randomness and enable the DM to focus on generating higher-likelihood trajectories from the dataset, we employ low-temperature sampling with a scaling factor $\xi \in \left[0,1\right)$\cite{ajay2022conditional}. This approach reduces the variance of the Gaussian distribution at each step of the denoising process, progressively lowering the noise intensity as the model iteratively refines the generated trajectory. By applying this scaling factor at every step, we prevent the degradation of model generation quality due to the contamination of suboptimal behaviors in the offline dataset, which helps stabilize the exploration process and ensures that the model generates more accurate and reliable trajectories. In addition, to ensure consistency between the generated trajectory and the historical data, the first state of the trajectory is always set to the currently observed state throughout all diffusion steps. After completing the final $K$ iterations of the denoising process, we can infer the next researchable predicted state  and determine the optimal action using our inverse dynamics model $f_{\phi}$. This procedure repeats in a standard receding-horizon control loop described in Algorithm ~\ref{implementing}.

\begin{algorithm}[!t]
    \caption{Implementation of CDMP \& CDMP-pen.}
    \label{implementing}
    \renewcommand{\algorithmicrequire}{\textbf{Input:}}
    \renewcommand{\algorithmicensure}{Initialize}
    \renewcommand{\algorithmiccomment}[1]{\hfill $\triangleright$ #1}
    
    \begin{algorithmic}[1]
        \REQUIRE Diffusion noise model $\epsilon_{\theta}$, inverse dynamics model $f_{\phi}$, planning horizon $H$, diffusion steps $K$, conditional guidance scale $\omega$, condition $\bm{y}$ 
        \ENSURE $t\gets0$
        
        \WHILE{not done}
            \STATE Observe $s_t$; Initialize $\bm{x}_K(\tau) \sim \mathcal{N}(\textbf{0},\xi \bm{I})$
            \STATE $\bm{x}_K(\tau)[0]\gets s_t$ \COMMENT{Constrain consistency}
            \FOR{$k=K\cdots1$}
                \STATE $\hat{\epsilon}\gets\epsilon_{\theta}(\bm{x}_k(\tau),k)+\omega(\epsilon_{\theta}(\bm{x}_k(\tau),\bm{y},k)-\epsilon_{\theta}(\bm{x}_k(\tau),k))$ \\ \COMMENT{Classifier-free guidance}
                \STATE $(\mu_{k-1}, \Sigma_{k-1})\gets{\rm{Denoise}}(\bm{x}_k(\tau),\hat{\epsilon})$
                \STATE $\bm{x}_{k-1}(\tau)\sim \mathcal{N}(\mu_{k-1}, \xi \Sigma_{k-1})$
                \STATE $\bm{x}_{k-1}(\tau)[0]\gets s_t$ \COMMENT{Constrain consistency}
            \ENDFOR
        \STATE Extract $(s_t, s_{t+1})$ from $\bm{x}_0(\tau)$
        \STATE Execute $a_t=f_{\phi}(s_t, s_{t+1})$; $t\gets t+1$
        \ENDWHILE
    
    \end{algorithmic}
\end{algorithm}

\section{Performance Evaluation}\label{V}
In this section, we conduct numerical experiments to evaluate the performance of our algorithm and validate the effectiveness of the proposed framework. Additionally, we explore the influence of different hyperparameters on the algorithm's behavior to further demonstrate its robustness.
\subsection{Experimental Settings}
We represent the noise model $\epsilon_{\theta}$ with a temporal U-Net\cite{ronneberger2015u}, consisting of $6$ repeated residual blocks. Timestep and condition embeddings are concatenated together and then added to the activations of the first temporal convolution within each block. The inverse dynamics model $f_{\phi}$ is developed as a self-regressive network generating $ML$-dimensional  actions, where each dimension is predicted sequentially based on the previous ones. 

We select OPNET as the simulation platform of a clustered wireless ad hoc network, where the environment spans a $10$ km $\times$ $10$ km area with randomly placed $16$ nodes. Each node has a maximum transmission range of $6$ km and can transmit packets in $4$ channels at a rate of $2$ Mbps. The traffic transmission stage of the frame consists of $10$ time slots. We simulate diverse network loads by adjusting the ratios of high- to low-speed nodes from $4:12$, $6:10$, to $8:8$. To collect the offline dataset, we allocate time slots based on these ratios to generate actions in multiple random communication contexts, recording environment data every frame. This yields a dataset of $1,814$ action-state trajectories labeled with rewards, each corresponding to a communication duration of $30$ seconds(i.e., $6,000$ time-steps).  

All algorithms including our proposed CDMP \& CDMP-pen, MFRL approaches such as Behavior Cloning (BC)\cite{farag2018behavior}, Conservative Q-learning (CQL)\cite{kumar2020conservative}, Implicit Q-learning (IQL)\cite{Kostrikov2021OfflineRL} and Decision Transformer (DT)\cite{chen2021decision}, as well as the MBRL method Diffuser\cite{janner2022planning}, are then trained on this offline dataset for $100$ epochs, with each epoch comprising of $1,000$ training steps. During testing, we evaluate the training planners on the aforementioned OPNET platform under random communication scenarios and compare performance based on the average reward and QoS (e.g., average throughput, average delay, and packet loss rate) of $5$-second episodes. We report the mean and the standard deviation over $3$ random scenarios. Furthermore, the principal parameters used in simulations are summarized in Table \ref{tab:parameters}.

\begin{table}[!t]
    \caption{List of Key Parameter Settings for the Simulation.}
    \label{tab:parameters}
    \centering
    \renewcommand{\arraystretch}{1.2}
    \begin{tabularx}{\linewidth}{|X|p{2cm}|} 
        \hline
        \textbf{Parameters Description} & \textbf{Value} \\
        \hline
        \multicolumn{2}{|l|}{\textbf{Simulator}} \\
        \hline
        Number of nodes & $N = 16$ \\
        Number of time slots in a frame & $M = 10$ \\
        Number of channels & $L = 4$ \\
        Coverage area of the simulation environment & $10$ km $\times$ $10$ km \\
        Maximum transmission range of each node & $6$ km \\
        Packet transmission rate & $2$ Mbps \\
        Ratios of high- to low-speed nodes & $4:12$, $6:10$, or $8:8$ \\
        Number of trajectories in the offline dataset & $1,814$ \\
        Communication duration of a trajectory in the offline dataset & $30$ s \\
        Communication duration of an episode during testing & $5$ s \\
        \hline
        \multicolumn{2}{|l|}{\textbf{CDMP \& CDMP-pen}} \\
        \hline
        Diffusion steps & $K = 200$ \\
        Planning horizon & $H = 12$ \\
        Discount factor & $\gamma = 0.99$ \\
        Conditional guidance scale & $\omega = 1.6$ \\
        Conditional information removal probability & $\beta = 0.25$ \\
        OOD penalty weight & $\zeta = 1.0$ \\
        \hline
        \multicolumn{2}{|l|}{\textbf{Neural Network Training}} \\
        \hline
        Architecture of the noise model $\epsilon_{\theta}$ & U-Net \\
        Architecture of the inverse dynamics model $f_{\phi}$ & Self-regressive network \\
        Dimension of diffusion step embedding in $\epsilon_{\theta}$ & $256$ \\
        Dimension of condition embedding in $\epsilon_{\theta}$ & $256$ \\
        Dimension of state embedding in $f_{\phi}$ & $256$ \\
        Dimension of action embedding in $f_{\phi}$ & $128$ \\
        Batch size & $128$ \\
        Number of training epochs & $N_{\rm{epoch}} = 100$ \\
        Number of training steps per epoch & $N_{\rm{step}} = 1,000$ \\
        Learning rate & $1e-4$ \\
        Optimizer & Adam \\
        EMA decay factor for model updates & $0.995$ \\
        \hline
    \end{tabularx}
\end{table}

\subsection{Performance Comparison}
\subsubsection{Numerical Results of CDMP} 
\begin{figure}[tbp]
\centerline{\includegraphics[width=\linewidth]{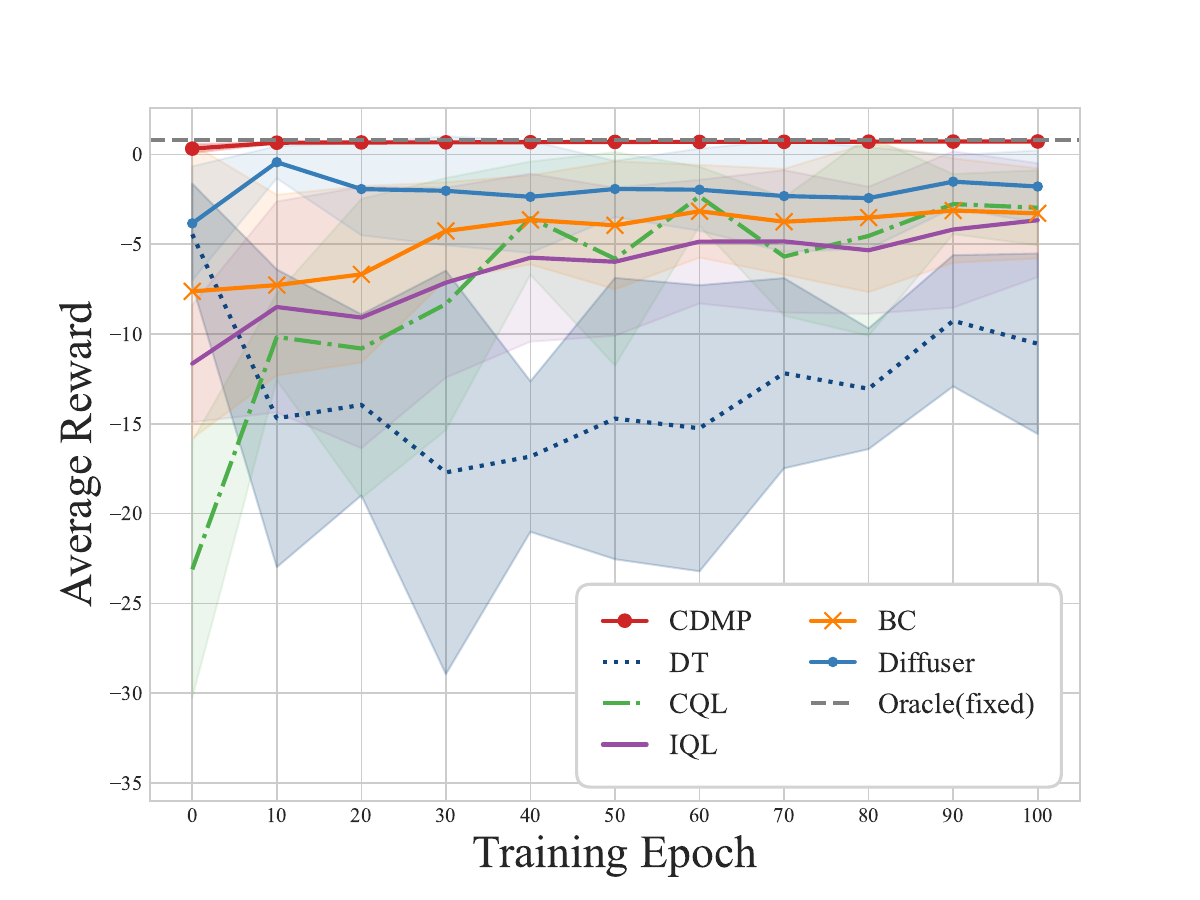}}
\vspace{-0.4cm}
\caption{Comparison of CDMP with different methods in terms of average reward.}
\label{fig3}
\vspace{-0.2cm}
\end{figure}

Firstly, we demonstrate the superiority of CDMP compared to other methods and plot the average reward curves during training, as shown in Fig.~\ref{fig3}. Notably, the oracle represents the optimal strategy, where the ratio of high- to low-speed nodes is known in advance, and time slots are allocated proportionally according to the ratio at the very beginning. It can be observed that CDMP achieves higher average rewards and approaches optimal performance after sufficient training. It also converges significantly faster with a smaller training variance, indicating a more stable training process.

\begin{figure}[t]
\centerline{\includegraphics[width=\linewidth]{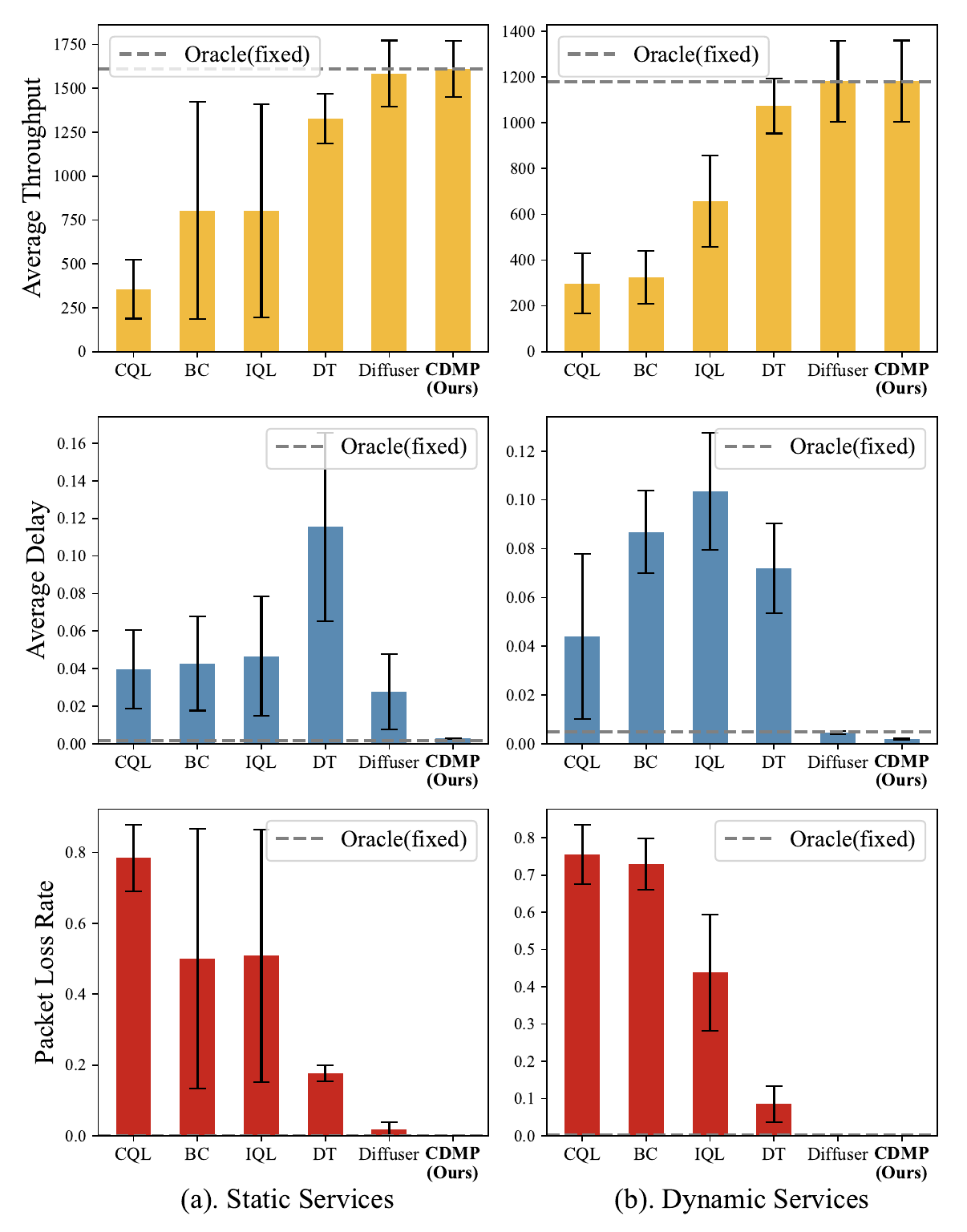}}
\vspace{-0.4cm}
\caption{Comparison of CDMP with different methods in terms of QoS. Specifically, static services represent a constant ratio of high- to low-speed nodes throughout the communication period. In contrast, dynamic services involve a changing ratio of high- to low-speed nodes over time.}
\label{qos}
\vspace{-0.2cm}
\end{figure}

Additionally, to further validate the effectiveness of CDMP, we compare the detailed QoS metrics, as shown in Fig.~\ref{qos}. The results demonstrate that CDMP outperforms other methods in terms of average throughput, average delay, and packet loss rate. Notably, in dynamic service scenarios, where the high- to low-speed nodes change along with time, CDMP even surpasses the oracle strategy, which remains fixed as time evolves. In contrast, CDMP can adaptively adjust time slot allocation, effectively accommodating these dynamic changes.

\subsubsection{Empirical Analysis of CDMP-pen}
\begin{figure}[tbp]
\centerline{\includegraphics[width=\linewidth]{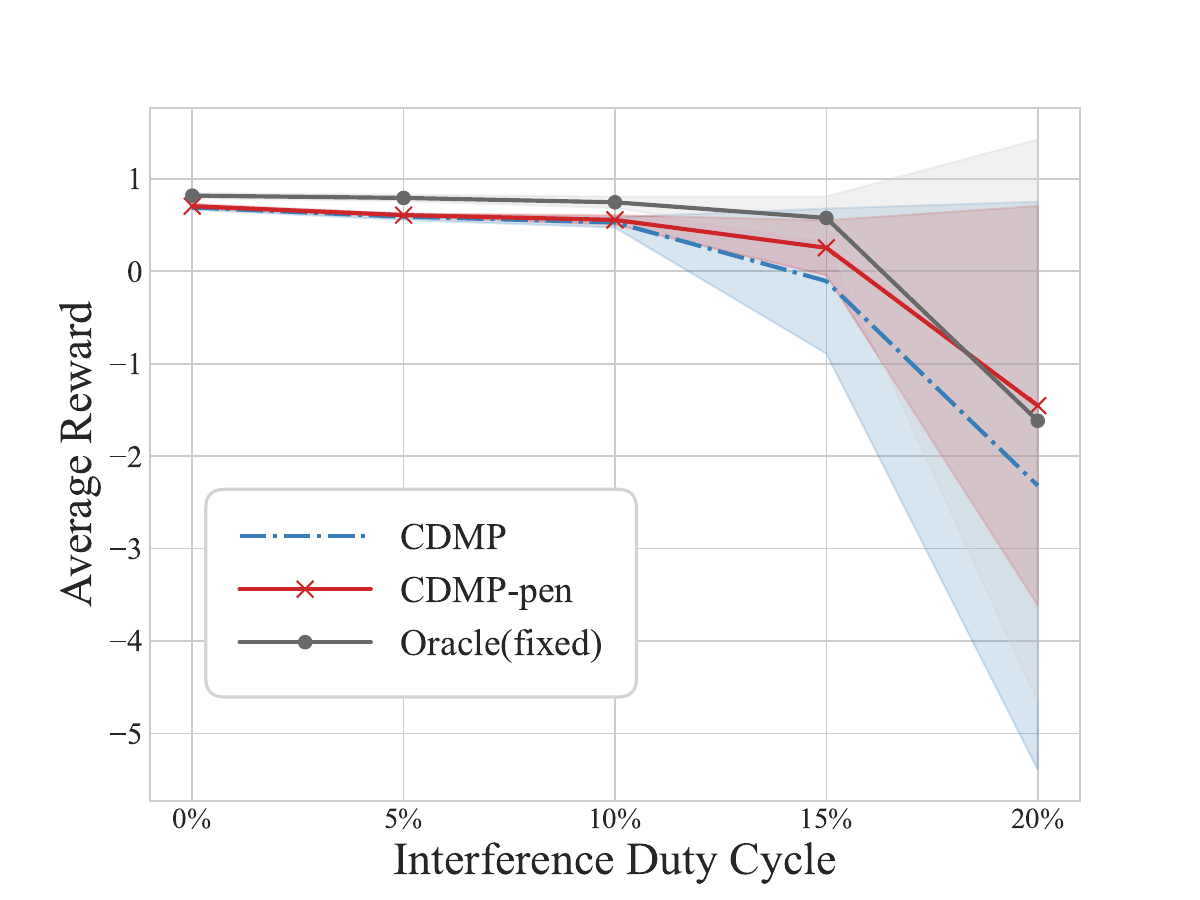}}
\vspace{-0.4cm}
\caption{Comparison of CDMP-pen and CDMP under different inference duty cycles in terms of average reward.}
\label{CDMP-pen}
\vspace{-0.2cm}
\end{figure}

To substantiate the potency of the proposed uncertainty penalty in addressing the distribution shift problem, we compare the performance of CDMP-pen and CDMP under varying interference duty cycles, as shown in Fig.~\ref{CDMP-pen}. In the experiment, OOD scenarios are simulated by introducing random ``interference'' packets into two channels. The comparison reveals that as the interference duty cycle increases, both CDMP and CDMP-pen experience performance degradation. However, CDMP-pen exhibits a slower decline. Moreover, under the same interference level, CDMP-pen achieves a higher average reward, indicating that the designed distance-to-data metric effectively quantifies the errors introduced by an imperfect dynamic model. This mechanism helps reduce model confidence in OOD regions, preventing high-risk decisions in unknown scenarios while enhancing policy robustness and generalization.

\subsection{Ablation Studies}
\begin{figure}[t]
\centerline{\includegraphics[width=\linewidth]{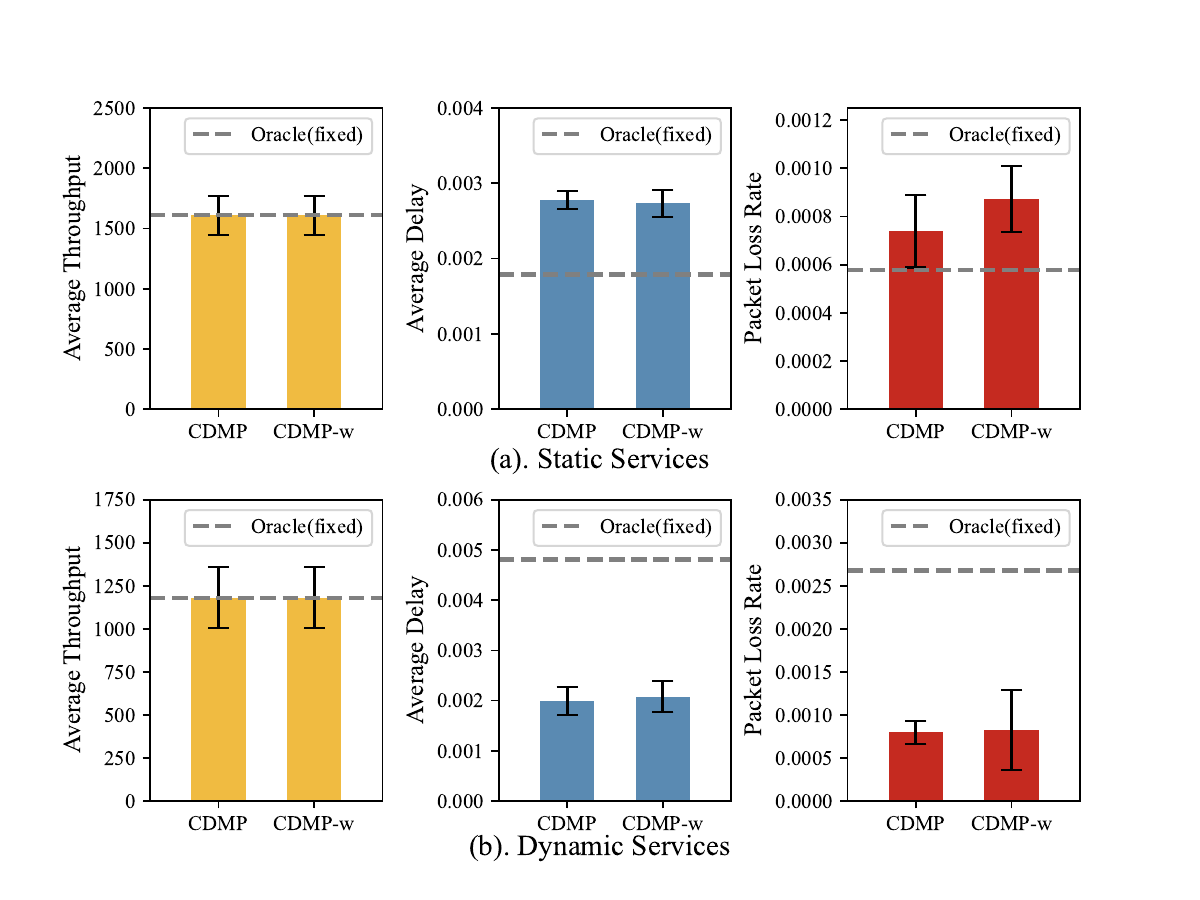}}
\vspace{-0.4cm}
\caption{Comparison of different reward function definitions in terms of QoS.}
\label{rewards}
\vspace{-0.4cm}
\end{figure}

To evaluate the efficacy and simplicity of the reward function defined in Eq. \eqref{eq:reward}, we design further experiments to investigate how different reward function definitions affect the algorithm's performance. Particularly, we re-define the QoS-weighted average reward function as
\begin{equation}
{\mathcal{R}}_{t}\mbox{-}{\rm{w}} :=\sum\nolimits_{i=1}^{N}\left ( \lambda \cdot u_{t}^{(i)} -  d_{t}^{(i)} \right )-\eta \cdot l_{t}, 
\end{equation}
where $u_{t}^{(i)}$ represents the size of data packets successfully transmitted and received, while $l_{t}$ denotes the ratio of lost packets to total sent packets in the network. The factors $\lambda$ and $\eta$ balance the weights of these metrics, aiming to promote high throughput while minimizing delay and packet loss across the entire communication system. The framework trained with this weighted reward is denoted as CDMP-w, and we compare the QoS metrics with CDMP as illustrated in Fig.~\ref{rewards}. As observed from the figure, using average delay alone as the reward function yields performance comparable to that of the QoS-weighted average reward function. Furthermore, it leads to a lower packet loss rate and a smaller standard deviation. This indicates that the simplified reward function not only circumvents the complexities of hyperparameter tuning but also contributes to improved model stability.

\begin{figure}[t]
\centerline{\includegraphics[width=0.9\linewidth]{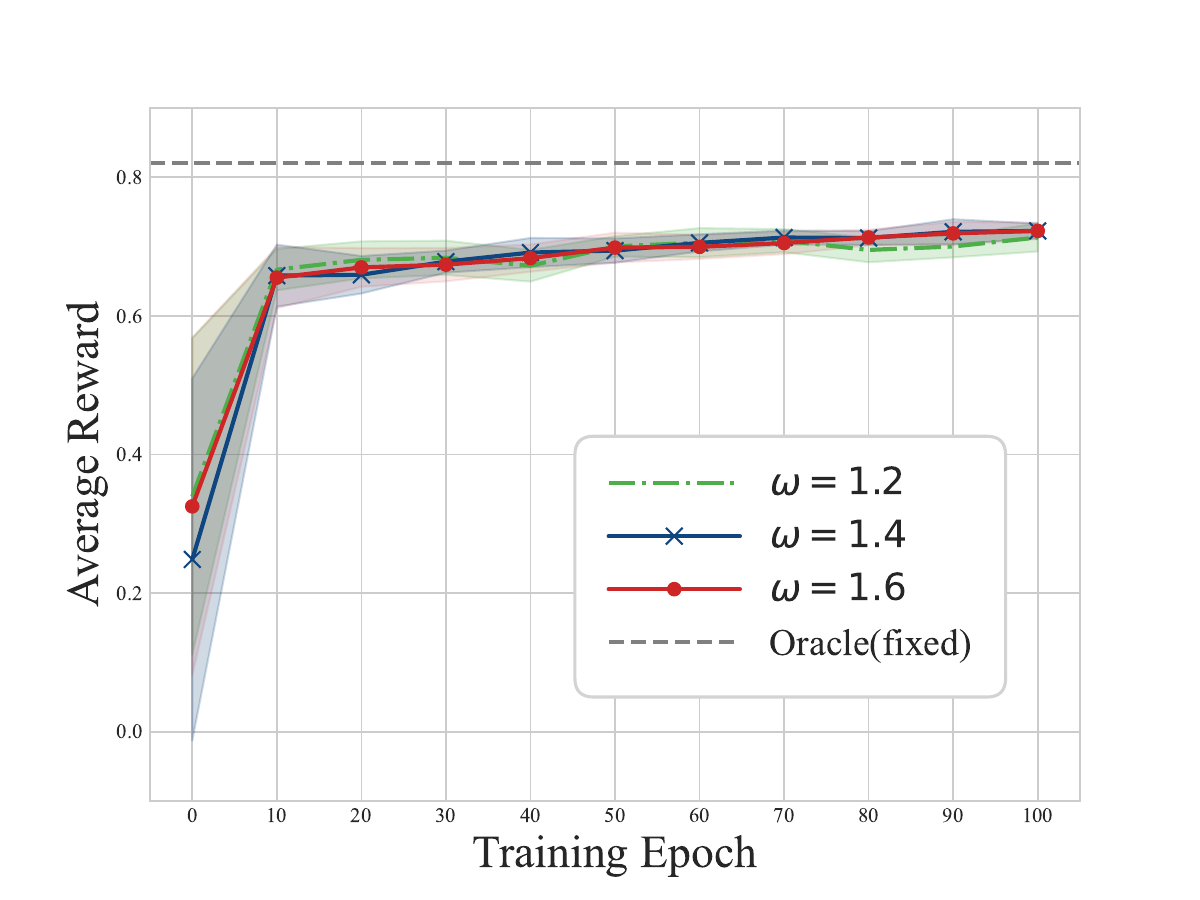}}
\vspace{-0.2cm}
\caption{Comparison of different conditional guidance scales $\omega$ in terms of average reward.}
\label{fig4}
\vspace{-0.4cm}
\end{figure}

Next, we investigate the impact of conditional guidance scale $\omega$ on the algorithm's performance of CDMP. By varying $\omega$, we track the average reward during training, as illustrated in Fig.~\ref{fig4}. The results indicate that for different $\omega$ settings, the model consistently converges to similar average reward values once completing sufficient training. This demonstrates the robustness and adaptability of the model with respect to variations in conditional guidance scalar, ensuring reliable results under varying conditions. However, small $\omega$ values may prevent the model from effectively capturing contextual cues, while excessively large $\omega$ values may cause the algorithm to rely too heavily on conditional guidance. Therefore, in practical implementation, the algorithm must strike a balance between stability and adaptability, carefully weighing the importance of conditional information.

\begin{figure}[tbp]
\centerline{\includegraphics[width=0.9\linewidth]{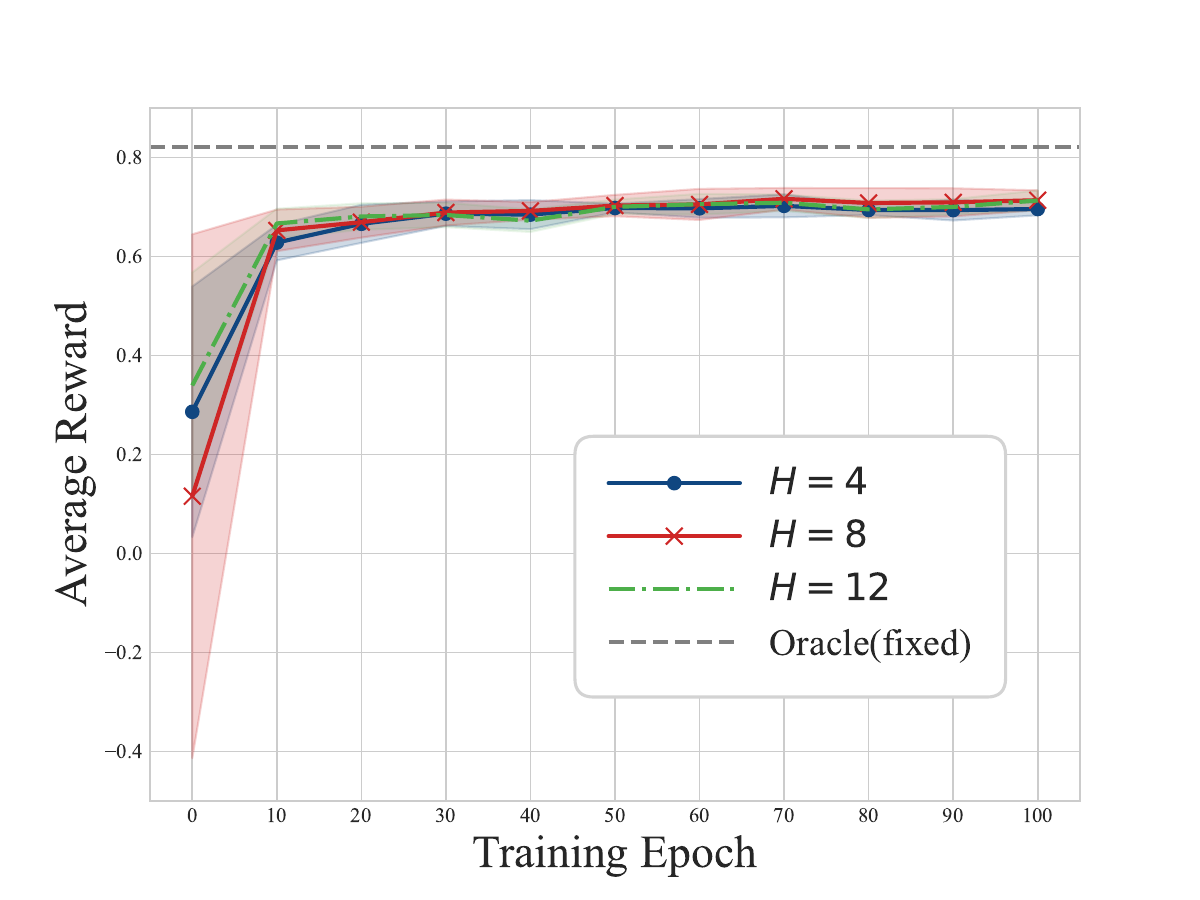}}
\vspace{-0.2cm}
\caption{Comparison of different planning horizons $H$ in terms of average reward.}
\label{fig5}
\vspace{-0.2cm}
\end{figure}

Furthermore, to verify the relationship between the planning horizon $H$ and the performance of CDMP, we conduct relevant experiments to evaluate how different values of $H$ affect average rewards. It can be observed from Fig.~\ref{fig5} that different planning horizons result in fluctuating model performance at the beginning of training. However, all settings rapidly converge to similar, high average rewards, indicating that the model maintains stable performance and strong reliability across a range of planning horizon configurations. Nonetheless, an excessively short horizon may neglect long-term rewards and future states, while longer horizons increase the computational complexity, impairing overall algorithm efficiency. Therefore, the planning horizon could be appropriately adjusted according to environmental complexity and available computational resources, so as to achieve an optimal balance between performance and efficiency.

\begin{figure}[tbp]
\centerline{\includegraphics[width=0.9\linewidth]{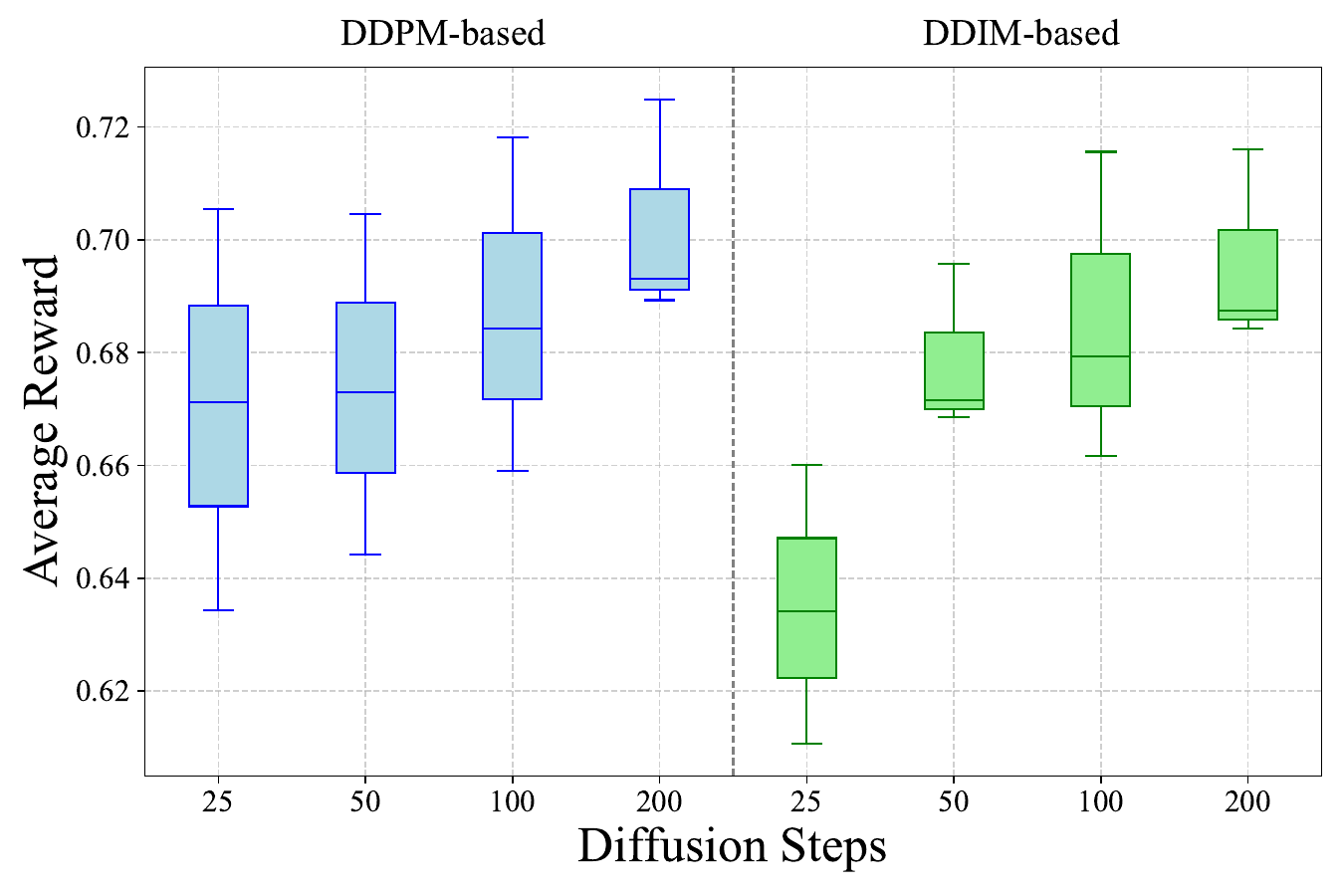}}
\vspace{-0.2cm}
\caption{Comparison of different diffusion steps $K$ and the denoising sampling mechanisms in terms of average reward.}
\label{K}
\vspace{-0.2cm}
\end{figure}

We also conduct a more in-depth numerical analysis on the effects of diffusion steps $K$ and the denoising sampling mechanism in CDMP, with results presented in Fig.~\ref{K}. The sampling process is built upon the DDPM and DDIM \cite{song2021denoising} frameworks, where DDIM leverages a pre-trained DDPM-based noise model ($K=200$) and follows a deterministic, non-Markovian strategy with uniform stride skipping to reduce iterations. Our comparison reveals that the DDPM-based model exhibits a steady increase in average reward as $K$ grows, with a concurrent reduction in standard deviation, enhancing performance robustness. In contrast, the DDIM-based model results in lower average rewards with smaller $K$, likely due to reduced denoising refinement and incomplete noise removal. Additionally, DDIM's deterministic nature causes convergence to similar solutions with fewer steps, resulting in lower standard deviation and decreased sample diversity. Nevertheless, reducing the number of iterative sampling steps accelerates generation and enhances computational efficiency. Moreover, DDIM’s training-free property enables seamless deployment of pre-trained models in real-world applications, lowering training costs while preserving competitive performance and adaptability to existing frameworks. Therefore, the choice of sampling framework and $K$ should be carefully tailored to the specific task, balancing the model's stability and complexity requirements. To our knowledge, this is the first study comparing DDPM and DDIM sampling frameworks in decision-making applications of DMs.

\begin{figure}[tbp]
\centerline{\includegraphics[width=0.9\linewidth]{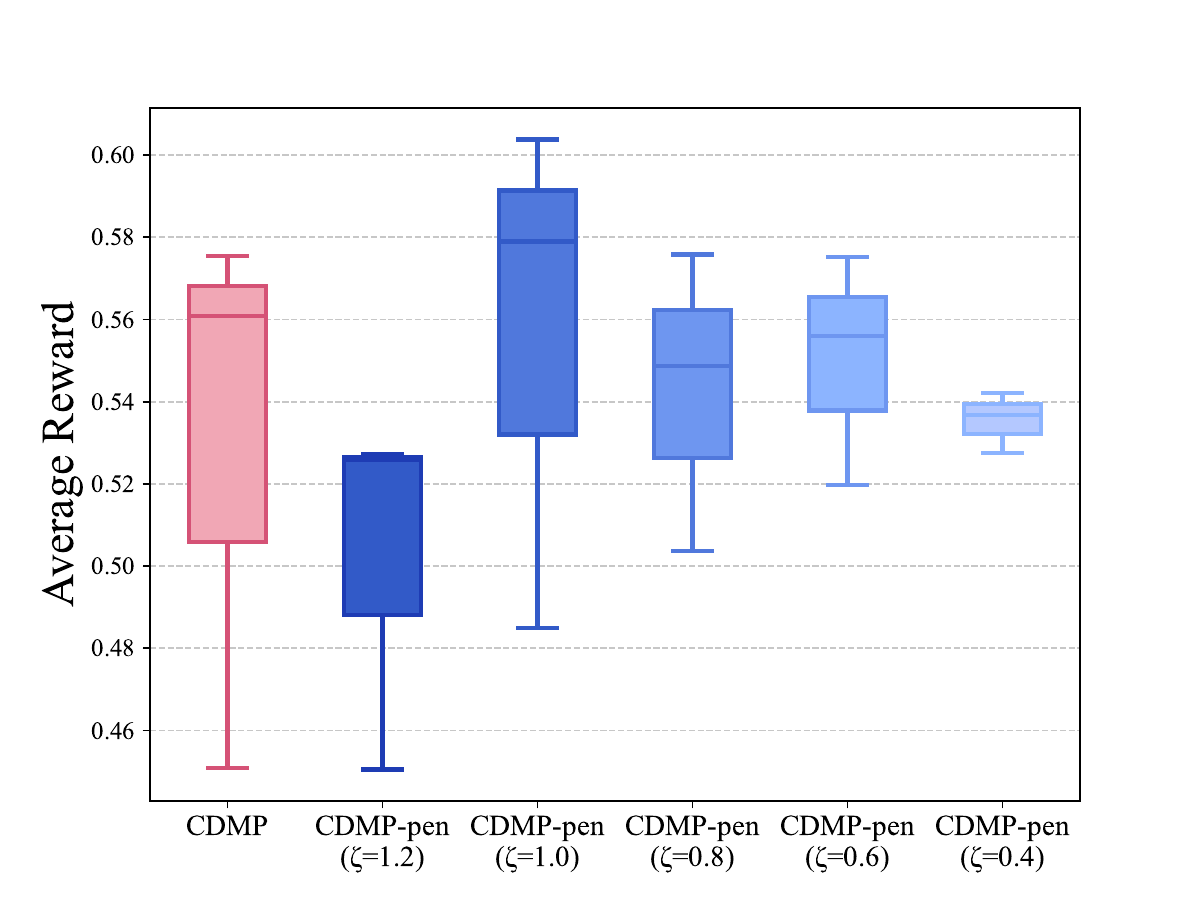}}
\vspace{-0.2cm}
\caption{Comparison of different OOD penalty weight factors $\zeta$ in terms of average reward.}
\label{zeta}
\vspace{-0.2cm}
\end{figure}

In addition, we investigate the impact of the OOD penalty weight factor $\zeta$ in the CDMP-pen loss function defined in Eq. \eqref{eq:CDMP-pen loss}. Experiments are conducted under a scenario with a $10\%$ interference duty cycle, where the average rewards of different weighting strategies are recorded, as illustrated in Fig.~\ref{zeta}. Comparative analysis reveals that CDMP-pen algorithms which impose penalties on unknown and hazardous states that deviate significantly from the offline dataset exhibit lower variance, demonstrating enhanced adaptability and robustness in dynamic environments. Moreover, appropriate parameter tuning enables the CDMP-pen algorithm to achieve a balance between conservatism and exploration, resulting in higher average rewards compared to CDMP.

\section{Conclusion}\label{VI}
In this paper, we have studied a centralized MBRL-based planning solution for MF-TDMA time slot scheduling strategy optimization. In particular, we have proposed the CDMP algorithm with a significant performance boost in terms of communication efficiency and effectiveness, addressing the challenges of resource allocation in clustered ad hoc networks. Specifically, we have utilized a conditional diffusion model to generate high-quality state sequences and employed an inverse dynamics model to predict the optimal actions. Afterward, we have introduced a dynamics uncertainty penalty to tackle the OOD problem in offline RL, extending it into a CDMP-pen algorithm, which has been theoretically proven to effectively bound the true distribution error. We have demonstrated the superiority \& robustness of our framework over several classical offline RL algorithms through extensive simulations.

There are many interesting directions to be addressed in the future. For example, our current simulation scenarios are limited to relatively simple and traditional MF-TDMA communication frameworks. Upon validating the proposed algorithm's effectiveness and advantages, future research could explore its integration with more sophisticated communication protocols. This would enable a more comprehensive investigation of the algorithm’s generalization capabilities and adaptability, broadening its practical applications and contributing to further advancements in communication systems. Moreover, our framework relies on centralized resource scheduling. 
Thus, as the network scales up, it significantly increases the complexity of training. Given these challenges, it is worthwhile to explore distributed decision-making approaches. Through localized task allocation and parallel training, distributed strategies can facilitate more efficient learning of optimal policies at the individual level, ultimately alleviating the burden of centralized information exchange and accelerating convergence.


%

\appendices

\section{Proof for Lemma \ref{lemma:equivalence}}\label{proof:lemma 1}
The perturbed data distribution can be written as a sum of Gaussians, since
\begin{align}
\hat{q}_\sigma(\widetilde{s}; \mathcal{D}_{\rm{stat}}) &:= \int \hat{q}(s; \mathcal{D}_{\rm{stat}}) \mathcal{N}(\widetilde{s}; s, \sigma^2 \mathbf{I}) ds \nonumber\\
&= \int \left[ \frac{1}{M'} \sum_i \delta(s^i) \right] \mathcal{N}(\widetilde{s}; s, \sigma^2 \mathbf{I}) ds \nonumber \\
&= \frac{1}{M'} \sum_i \int \delta(s^i) \mathcal{N}(\widetilde{s}; s, \sigma^2 \mathbf{I}) ds\nonumber\\
&= \frac{1}{M'} \sum_i \mathcal{N}(\widetilde{s}; s^i, \sigma^2 \mathbf{I}). 
\end{align}
Then we consider the negative log of the perturbed data distribution multiplied by $\sigma^2$ as
\begin{align}
-\sigma^2& \log \hat{q}_\sigma(\widetilde{s}; \mathcal{D}_{\rm{stat}})\! = \!-\sigma^2 \log \left[ \frac{1}{M'}\! \sum_i \mathcal{N}(\widetilde{s}; s^i, \sigma^2 \mathbf{I}) \right] \nonumber\\
&\!=\! -\sigma^2 \log \left[ \sum_i \mathcal{N}(\widetilde{s}; s^i, \sigma^2 \mathbf{I}) \right] + \sigma^2 \log M' \nonumber\\
&\textstyle\!=\! -\sigma^2\! \log\! \left[ \frac{1}{\sqrt{(2\pi\sigma)^n}} \sum_i \exp\! \left( \!-\!\frac{1}{2\sigma^2} \|\widetilde{s}\! -\! s^i\|^2 \right) \right]\! +\! \sigma^2 \log M' \nonumber\\
&\!=\! -\sigma^2 \log \left[ \sum_i\! \exp\! \left( -\frac{1}{2\sigma^2} \|\widetilde{s} \!-\! s^i\|^2 \right) \right]\! \nonumber\\
&\hspace{2.9em}+\! \sigma^2 \log M'\! +\! \frac{\sigma^2 n}{2} \log(2\pi\sigma) \nonumber\\
&\!=\! -\sigma^2 \text{LogSumExp}_i \left[ -\frac{1}{2\sigma^2} \|\widetilde{s} - s^i\|^2 \right] + C(M', n, \sigma) \nonumber\\
&\!=\! \text{Softmin}_\sigma \left[ \frac{1}{2} \|\widetilde{s} - s^i\|^2 \right] + C(M', n, \sigma)\nonumber \\
&\!=\! d_\sigma(\widetilde{s}; \mathcal{D}_{\rm{stat}})^2+C(M', n, \sigma).
\end{align}
Then the lemma comes.
\hfill $\blacksquare$

\section{Proof for Theorem \ref{theorem:bound}}\label{proof:lemma 2}
Let $L_e$ denote the local Lipschitz constant of the true error $e(s)$ valid over the state space $\mathcal{S}$, and define $s_c := \mathop{\arg\min}\limits_{s^i \in \mathcal{D}_{\rm{stat}}}\frac{1}{2} {\parallel s - s^i \parallel}_2^2$, i.e., the closest data-point. By definition, we have $e(s) \leq e(s^i) + L_e \|s - s^i\|_2 $. Consequently, the following holds:
\begin{align}
e(s) &\leq \min_{s^i \in \mathcal{D}_{\rm{stat}}} \left[ e(s^i) + L_e \|s - s^i\|_2 \right] \nonumber\\
&\leq e(s_c) + L_e \|s - s_c\|_2 \nonumber\\
&= e(s_c) + \sqrt{2}L_e \sqrt{\frac{1}{2} \|s - s_c\|_2^2} \nonumber\\
&= e(s_c) + \sqrt{2}L_e \sqrt{\frac{1}{2} \min_{s^i \in \mathcal{D}_{\rm{stat}}} \|s - s^i\|_2^2} \nonumber\\
&\textstyle \leq e(s_c)\! +\! \sqrt{2}L_e \sqrt{-\sigma^2 \log \! \left(\frac{1}{M'} \sum_i \exp \!\left( -\frac{1}{2\sigma^2} \|s \!-\! s^i\|_2^2 \right) \right)} \nonumber\\
&= e(s_c) + \sqrt{2}L_e \sqrt{d_\sigma(s; \mathcal{D}_{\rm{stat}})^2 + C_2},
\end{align}
where $C_2 = \sigma^2 \log M' - C_1$, with $C_1$ defined in Eq. \eqref{eq:smoothed distance to data}. In the second line, we leverage the fact that as $s_c$ is a feasible solution to the minimization problem in the first line, it serves as an upper bound for the optimal value. In the fifth line, we rely on convex analysis and the characteristics of the smoothed minimum, where for any vector $\mathbf{v} = [v_1, \cdots, v_n]^\top \in \mathbb{R}^n$, the inequality $\min\{v_1, \cdots, v_n\} \leq -\frac{1}{\delta} \log \sum_{i=1}^n \exp(-\delta v_i) + \frac{\log n}{\delta}$ holds for any scaling parameter $\delta > 0$. 
Thus, we attain the theorem. \hfill $\blacksquare$



\end{document}